# Highly emissive, selective and omnidirectional thermal emitters mediated by machine learning for ultrahigh performance passive radiative cooling

Yinan Zhang[1#*], Yinggang Chen[1,2#], Tong Wang[1 #], Qian Zhu[1,2], & Min Gu[1 *]

[1]Institute of Photonic Chips, University of Shanghai for Science and Technology, Shanghai 200093, China.

[2]Centre for Artificial-Intelligence Nanophotonics, School of Optical-Electrical and Computer Engineering, University of Shanghai for Science and Technology, Shanghai 200093, China.

*Corresponding authors. Emails: zhangyinan@usst.edu.cn; gumin@usst.edu.cn

**Abstract**

Real-world passive radiative cooling requires highly emissive, selective, and omnidirectional thermal emitters to maintain the radiative cooler at a certain temperature below the ambient temperature while maximizing the net cooling power. Despite various selective thermal emitters have been demonstrated, it is still challenging to achieve these conditions simultaneously because of the extreme complexity of controlling thermal emission of photonic structures in multidimension. Here we demonstrated machine learning mediated hybrid metasurface thermal emitters with a high emissivity of ~0.92 within the atmospheric transparency window 8-13 μm, a large spectral selectivity of ~1.8 and a wide emission angle up to 80 degrees, simultaneously. This selective and omnidirectional thermal emitter has led to a new record of temperature reduction as large as ~15.4 °C under strong solar irradiation of ~800 W/m², significantly surpassing the state-of-the-art results. The designed structures also show great potential in tackling the urban heat island effect, with modelling results suggesting a large energy saving and deployment area reduction. This research will make significant impact on passive radiative cooling, thermal energy photonics and tackling global climate change.

**Introduction**

The accelerated global warming and environmental threats on human life have motivated enormous research and development on zero-carbon cooling technologies to replace or complement the conventional compressed air-based cooling technologies. Daytime passive radiative cooling is a recently demonstrated disruptive technology that could potentially transform the global energy landscape because of its zero-energy consumption and zero-carbon emission[1-6]. The success of this technology replies on the spectral control over both the solar and thermal wavelengths by advanced photonic structures and micro/nano-fabrication technologies that enable simultaneous reflection of strong solar irradiation and radiation of heat to the ultracold outer space [5-28].

Ideal selective thermal emitters with a unity emissivity in the atmospheric transparency window 8-13 μm and zero emissivity outside the window are highly demanded as they suppress the absorption from downward atmosphere thermal emission outside the transparency window, allowing much lower equilibrium temperature[2,29]. Although it has been previously demonstrated by a few simple and well-established photonic structures, including multilayer dielectric thin films[1,13-16], silica sphere/polymer metamaterials[10-12], plasmonic metasurfaces/nanoparticles[18-20], which utilize photonic resonances to enhance their infrared emission, the emission profiles of these structures are still highly dispersive with either low emissivity or mismatched with the transparency window. To achieve ideal selective thermal emissions, complex photonic structures with multiple geometry degrees of freedom and multiple photonic resonances and resonance couplings are highly required to mitigate the intrinsically highly dispersive emission nature; however, such a complexity has remained a challenging endeavor because of the elusive light-matter interactions at the micro/nano-scale.

On the other hand, the angular profile of the emissivity determines both the lowest steady temperature and the cooling power. In practical scenarios, maintaining the radiative coolers below the ambient temperature while maximizing the net cooling power is of critical importance because the radiative coolers exchange significant heat with surrounding environment through the non-radiative heat conductive and convection processes, such as the heat conduction by physically contacting with the experimental apparatus or the targeted cooling objects[4]. To achieve this, one needs to balance these external heating loads by widening the thermal emission angle in addition to the spectrally selective bandwidth, i.e., achieve selective and omnidirectional thermal emitters. However, very few studies have paid attention to the nontrivial emission angle issue, let alone achieving highly selective and wide emission, simultaneously.

To tackle the aforementioned challenges, we propose machine learning mediated hybrid metasurface thermal emitters for highly selective, strongly emissive, and omnidirectional thermal emission. Hybrid metasurfaces[30-32], engineered planar structures with complex building blocks featuring multiple physical degrees of freedom and photonic resonances, hold great promise in controlling the thermal emission dispersion. Machine learning, especially the multilayer perception neuron network, allows effectively and time-efficiently inverse designing photonic structures with targeted optical responses[33-37], enabling achieving complex photonic structures with previously unattainable functionalities and performance. By combining these two, we demonstrated thermal emitters with an emissivity of ~0.92 across the atmospheric transparency window, ultra-high spectral selectivity of ~1.8 and emission angle up to 80 degrees simultaneously, leading to a new record of equilibrium temperature reduction as large as ~15.4 °C in rooftop test, significantly surpassing the state-of-the-art radiative coolers under non-vacuum conditions. The demonstrated radiative cooler also holds great promise for the urban heat island mitigation with modeling results suggesting a more than 50% deployment area reduction compared with the current radiative coolers.

**Machine learning mediated hybrid metasurface thermal emitters**

The hybrid metasurface structure and machine learning for inverse design are shown in Figs. 1a-1c. The unit of the metasurface consists of alternative $SiO_2$ and $Si_3N_4$ micro-disks, forming a hybrid dielectric Mie resonator, which is mainly responsible for enhancing the thermal emission (Fig. 1a). The bottom layer of the structure is an optically thick Ag back reflector, which can reflect the incoming solar irradiation and enhance the light-mater interactions at the thermal wavelengths. The constituted material of the metasurface unit can also be other dielectric polar materials such as $Al_2O_3$, $TiO_2$, $HfO_2$ with phonon polariton resonances within the transparency window. In such a structure, a multitude of photonic resonances can exist, making the thermal dispersion control extremely challenging. Fig. 1d illustrates the electric field distributions at the resonance wavelengths of an optimized $SiO_2/Si_3N_4/SiO_2$ 3-layer geometry. Notably, photonic mixing resonances, including dominated Mie resonances at the wavelengths of 8.97 μm and 11.12 μm, gap resonances at 10.32 μm, and diffractive optical resonances at 7.7 μm, are observed, in addition to the phonon polariton resonances at the wavelengths of 9.3 μm and 11.5 μm (Fig. S1). Some of the resonances within the structure interfere with each other forming a hybridized electric field distribution because of the geometrical contact between $SiO_2$ and $Si_3N_4$ layers.

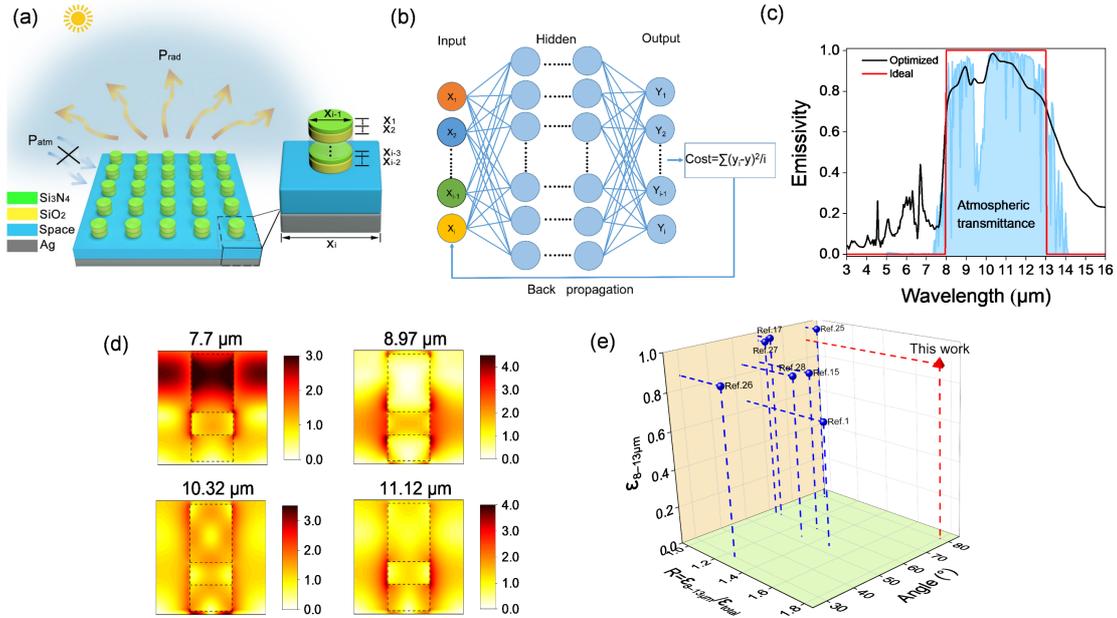

Fig. 1 Hybrid metasurface thermal emitter mediated by machine learning. (a) The proposed hybrid metasurface with the unit consisting of alternative $Si_3N_4$ and $SiO_2$ layers. $P_{atm}$ is the downward atmosphere thermal emission and $P_{rad}$ is the thermal emission of the metasurfaces. (b) Schematic of the multilayer perception neural network for inverse design with input, hidden and output layers. (c) Emissivity spectrum of the inversely optimized 3-layer selective thermal emitter, compared with the ideal spectrum. (d) The electrical field distributions at various resonance wavelengths of the optimized 3-layer ($SiO_2$/$Si_3N_4$/$SiO_2$) structures. (e) Emissivity, selectivity, and emission angle of this work, in comparison with state-of-the-art results.

The neuron network employed is a fully connected 4-layer perception neural network (Fig. 1b). The input is the geometry parameters of the structure, including the thickness of each layer, the unit diameter and the periodicity. The output is the spectrum sampled at points between 3 and 16 μm. Firstly, the neuron network is trained to forwardly predict the thermal emission spectra from a given set of structure parameters by the sample data obtained from the finite difference time domain calculations. Then the weights of the neuron network are fixed, and the back propagation method is used to train the input structure parameters to find the geometry that can mostly produce the ideal selective thermal emission. The training error graphed in Fig. S2a and cross-validation responses for various layer configurations in Table 1 confirm that the neuron network is well established and trained. Furthermore, the tested forward structure-spectra prediction and inverse spectra-structure design shown in Figs. S2b-c demonstrated the accuracy of the network in approximating the simulation and optimizing the structure. Thermal emitters with different layer numbers (2, 3 &4) are optimized, demonstrating highly emissive and selective thermal emissions in all cases (Fig. S3). Interestingly, even a 2-layer configuration can achieve excellent selective thermal emission, greatly easing the practical fabrication complexity. The optimized selective thermal emission spectra for a 3-layer structure are shown in Fig. 1c, suggesting an average thermal emissivity of ~0.92 across the atmospheric transparency window 8-13 μm and a spectral selectivity of ~1.8. More exhilaratingly, the high emissivity and selectivity was found to remain unchanged within an emission angle up to 80 degrees. This is so far the first thermal emitter with simultaneously high emissivity, high selectivity, and large emission angle to the best of our knowledge.

**The fabricated thermal emitter and infrared emissivity**

To experimentally demonstrate the high performance selective thermal emission, the machine-learned 2-layer hybrid metasurfaces are fabricated on top of a 4 inches silicon wafer with the bottom side coated with an optically thick Ag mirror, with Fig. 2a showing a photography of the fabricated sample (see Methods for details of the fabrication procedures). Fig. 2b shows the top-view and tilted SEM images of an array of the hybrid $SiO_2/Si_3N_4$ dielectric resonators, demonstrating a uniform 2D array and clear 2-layer structures. The measured diameter and periodicity of the structures are 4.22 μm and 9.15 μm, respectively. While the thickness of the $SiO_2$ and $Si_3N_4$ are 6.2 μm and 1.4 μm. The energy-dispersive X-ray spectroscopy elemental mappings of Si, O and N in Fig. 2c confirms the constitute material of the structure. We measured the emissivity spectra of the structure via Fourier-transform infrared (FTIR) spectroscopy with an unpolarized incident light beam, with the results shown in Fig. 2d. Clearly, the structure exhibits high-emissivity and high-selectivity emission profile, agreeing very well with the theoretically optimized results in Fig. S3a. The difference between the experimental and theoretical results are due to the imperfect fabrication and the minor dielectric constant difference. The measured average emissivity can achieve ~0.92 in the transparency window and ~0.6 among the entire range 3-16 μm, with a spectral selectivity of ~1.53. The relative smooth emission profiles in the transparency window are attributed to the multipolar Mie resonances and the material intrinsic absorption properties. To further reveal this, we performed a multipolar decomposition of the 2-layer dielectric Mie resonator. Two feature peaks at 8.9 μm and 10.5 μm appear in the spectrum, corresponding to the electric and magnetic dipole resonances, respectively. The electric and magnetic field profiles further confirm the magnetic and electric dipole resonances. The resonance enhances the electric field at the wavelengths where the material absorption is relatively weak and accordingly increase the total absorption/emission. For example, the electric field at the wavelength of 10.5 μm mainly concentrates at the bottom $SiO_2$ layer, enhancing the absorption/emission of $SiO_2$ at this wavelength.

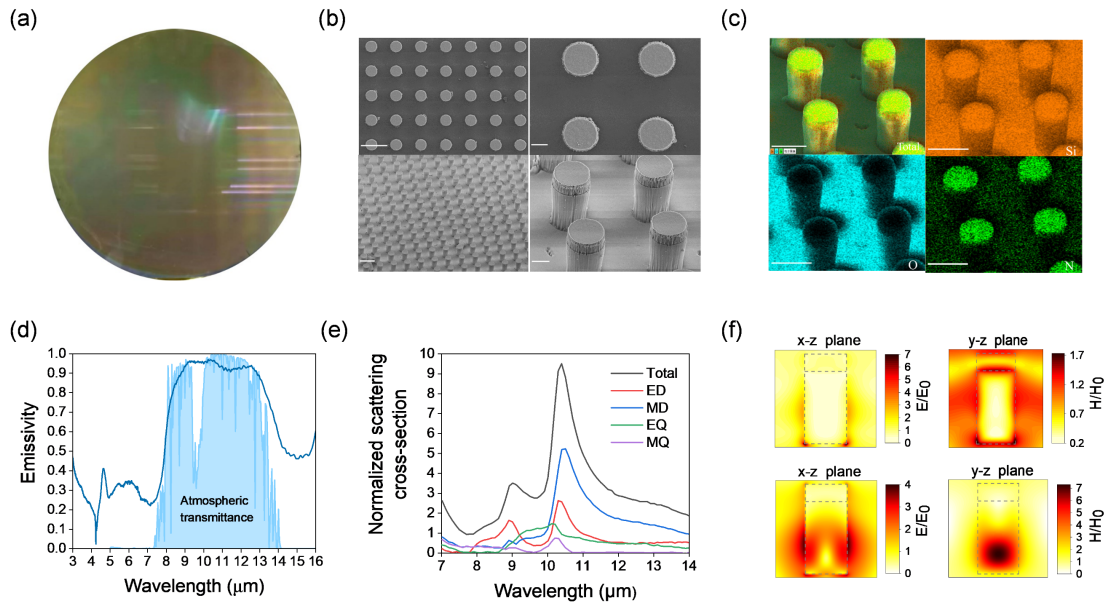

Fig. 2 The fabricated 2-layer metasurface and its emissivity. (a) A photography of the fabricated 4 inches metasurface thermal emitter. (b) Top-view and tilted SEM images of an array of the hybrid $SiO_2/Si_3N_4$ dielectric resonators (scale bar: 10 μm and 2 μm for the left and right two figures, respectively). (c) Energy-dispersive X-ray spectroscopy elemental mappings of the metasurfaces. (d) The measured infrared emissivity of the metasurfaces. (e) Multipolar decomposition of the 2-layer dielectric resonator. ED: electrical dipole; MD: magnetic dipole; EQ: electric quadrupole; MQ: magnetic quadrupole. (f) Electric and magnetic field profiles at the two resonance wavelengths 8.9 μm (top) and 10.5 μm (bottom).

**Angle-resolved thermal emission**

Apart from the spectral domain, the emission profile in angular domain also plays crucial roles. Generally, angular selective thermal emitter with an emission angle restricted in the zenith direction is believed to allow much lower equilibrium temperature because it avoids the oblique thermal radiation from the atmosphere. Indeed, the calculated net cooling power (Fig. 3a) of the radiative cooler under zero solar absorption and zero non-radiative heat exchange, indicating an optimum emission angle in the zenith direction for lower steady temperature (see Supplementary Notes on the detailed calculations). However, the real-world scenarios and applications involves unavoidable non-radiative heat exchange, i.e., heat conduction and convection, requiring larger cooling power to balance these external heating loads and might pose different angular constraints. Therefore, the emissivity profile needs to be reconsidered and reanalyzed in real-world applications. A theoretical calculation of the cooling power as a function of the temperature considering the non-radiative heat exchange ($h=2$ $W/m^2/K$) was performed with the results shown in Fig. 3b. It is demonstrated that wide-angle selective thermal emitter enables larger cooling power and lower equilibrium temperature as well under realistic conditions even with a small heat exchange coefficient of 2 $W/m^2/K$ (approximately the lowest value for non-vacuum apparatus). The steady temperature was also calculated with various emission angles under different non-radiative heat exchange coefficients (Fig. 3c), confirming the superiority of selective and omnidirectional thermal emitters in achieving lower steady temperature in practical scenarios.

Our selective thermal emitters show surprisingly high-performance omnidirectional emission properties, with the unpolarized emissivity under different angles shown in Fig. 3d. The selectivity can even preserve at an angle up to 80 degrees without obvious emission reduction across the transparency window, except a minor reduction within 40-60 degrees. This can be attributed to the relatively weak absorption/emission of s polarized light at the wavelengths of 11-13 μm (Fig. 3e) because of the reduced electric field distribution. Nevertheless, this does not influence the average thermal emission too much because of the highly emissive and selective spectral profile of the p polarized thermal emission (Fig. 3f).

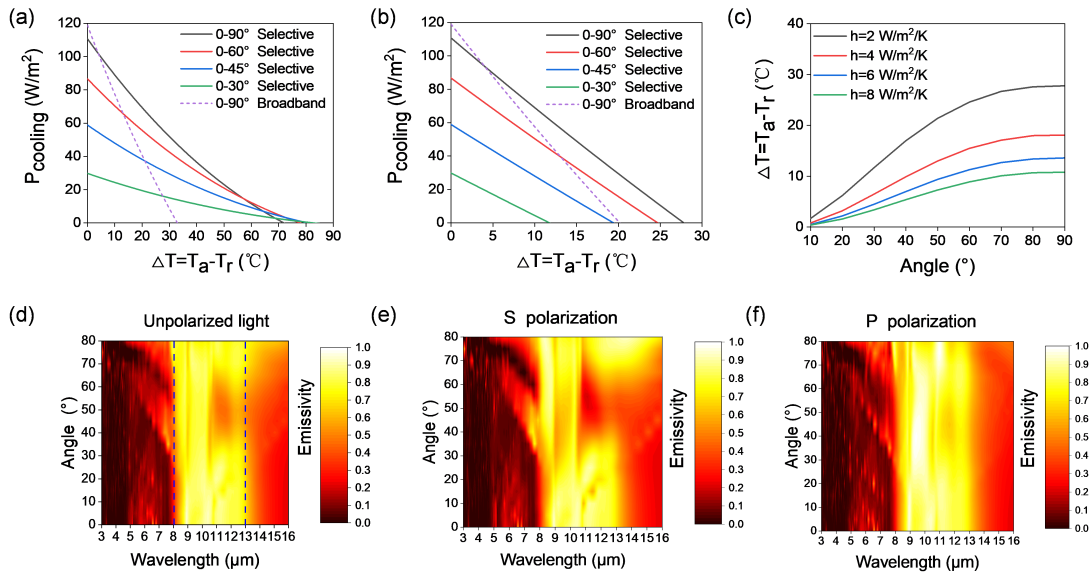

Fig. 3 Angle-resolved thermal emissivity. (a, b) Net cooling power $P_{cooling}$ of the ideal selective thermal emitter with various thermal emission angles in the cases of heat exchange coefficient $h =$ 0 W/m$^2$/K (a) and 2 W/m$^2$/K (b). $T_a$ and $T_r$ are ambient and radiative cooler temperature, respectively. (c) The steady temperature of the radiative cooler as a function of the emission angle for various heat exchange coefficients h. (d-f) The angle-resolved emissivity (unpolarized, s polarized, and p polarized) of the metasurface thermal emitter.

**Cooling performance of the radiative cooler**

To experimentally test the radiative cooling performance of the selective and omnidirectional thermal emitters, a continuous outdoor measurement of the temperature was performed on clear days in Shanghai, China. The schematic diagram of the outdoor apparatus is shown in Fig. 4a, where the sample is placed inside a foam with low thermal conductivity to minimize the non-radiative heat exchange. Al foil was used to wrap the foam such that the solar irradiation and surrounding thermal emission is reflected to avoid heating up the foam. On top of the setup, an IR transparent low density PE film was used to seal the sample preventing wind effect during the measurement. To increase the solar reflection, a home-made nanoporous PE film was integrated on top of the sample. Fig. 3b presents a real photo of the apparatus, where the pyranometer, the thermocouple, wind and humidity monitors are included. The solar reflection spectra of the nanoporous PE film and its infrared transmission are shown in Fig. S4, respectively. The integrated device shows high solar reflection across the entire solar wavelength range 0.3-2.5 μm, with an average solar reflectivity of ~95%. We performed the test on both winter and summer days. On the winter day, an average temperature reduction of 10 °C was realized, with a maximum temperature reduction of 14.7 °C when the sample was exposed to a solar irradiation of 600-650 W/m$^2$ at noon time (Fig. S5). On the summer day, a higher average temperature reduction of 12.7 °C with maximum reduction of 15.4 °C was demonstrated even the sample was exposed to much stronger solar irradiation of 800-850 W/m$^2$ (Figs. 4c-e). The demonstrated experimental results are obviously superior to state-of-the-art results with a temperature reduction less than ~10 °C. Considering the potential improvement room of the solar reflection, the temperature reduction can be further expected. To further elucidate the cooling performance of the fabricated thermal emitter, we calculated the cooling power as a function of the temperature reduction using the experimental measured solar and emission spectra. The heat exchange coefficient between the thermal emitter and the surrounding environment was calculated by the equation $h=2.5+2v$, where $v$ is the measured wind speed[38]. It is revealed the experimental maximum temperature reduction is close to the theoretical maximum value 16.5 °C, demonstrating excellent performance of our metasurface radiative coolers.

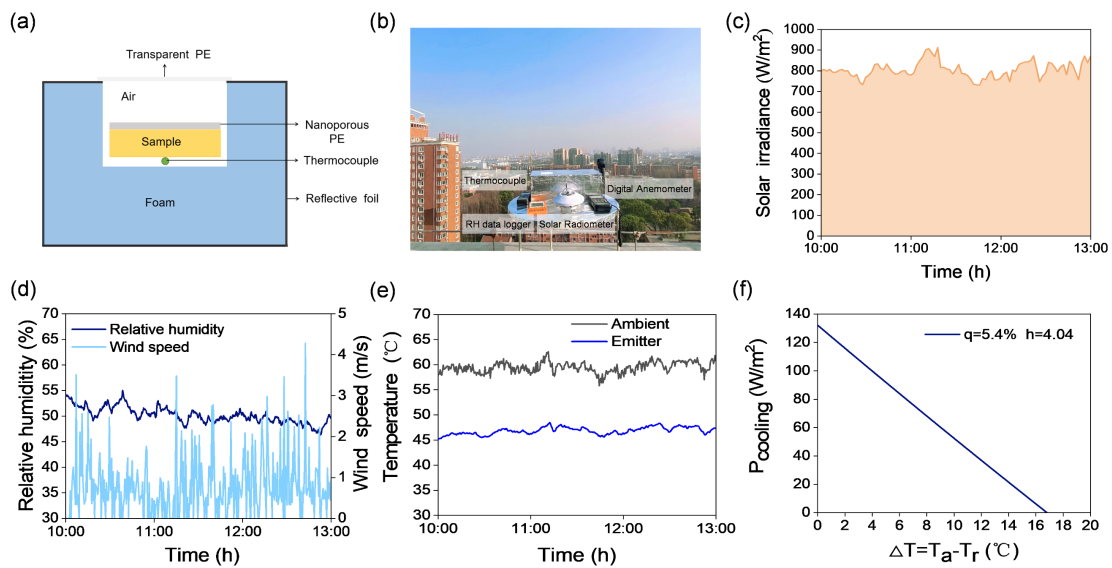

Fig. 4 (a) Schematic of the apparatus and the metasurface radiative cooler. (b) Photo of the apparatus on the test rooftop in Shanghai, China. (c, d) The solar irradiation and environment (wind speed and relative humidity) during the rooftop test. (e) Steady-state temperature of the metasurface radiative cooler and the ambient temperature. (f) The net cooling power of the radiative cooler using the experimental spectra data (solar absorbance: $q$=5.4%; heat exchange coefficient: $h$=4.04 W/m$^2$/K).

**Discussion**

The demonstrated selective and omnidirectional thermal emitters allow both lower steady temperature and larger cooling power, which is of critical importance for many practical cooling applications with nontrivial external heating load, such as the urban heat island effect. The urbanized areas usually experience higher temperatures up to 6 °C than outlying areas because of the net heat gain in the city areas[39]. Therefore, deploying radiative coolers in city areas can offset this net heat gain thereby preventing or even eliminating the urban heat island effect (Fig. 5a). The potential energy savings and deployment areas can be calculated. Fig. S6 shows the cooling power of our radiative cooler when exposed to an ambient temperature of 307 K (average summer temperature of Shanghai, China[40]) under various angle restriction using the simulated emission spectra. The required area for eliminating the net heat gain by using our wide-angle radiative cooler can be calculated (Fig. 5b). As can be seen, the deployment area by using our radiative cooler is significantly lower than that by current reported radiative coolers with an emission angle of ~60 degree. The area required to eliminate the urban heat island effect in typical summertime for 4 big cities in China: Beijing, Shanghai, Chongqing, Tianjin was further calculated using the reported net heat gain data (Figs. 5c-f)[41]. The results demonstrate that our wide-angle radiative cooler significantly reduce the deployment area by 35%. Furthermore, when reducing the temperature to a comfortable body temperature of 24 °C, the deployment area can be reduced by more than 50%. With the development of scalable manufacturing technologies of metasurfaces, it is believed the designed structures will bring broad and significant impact on relieving the urban heat island effect and other global warming issues.

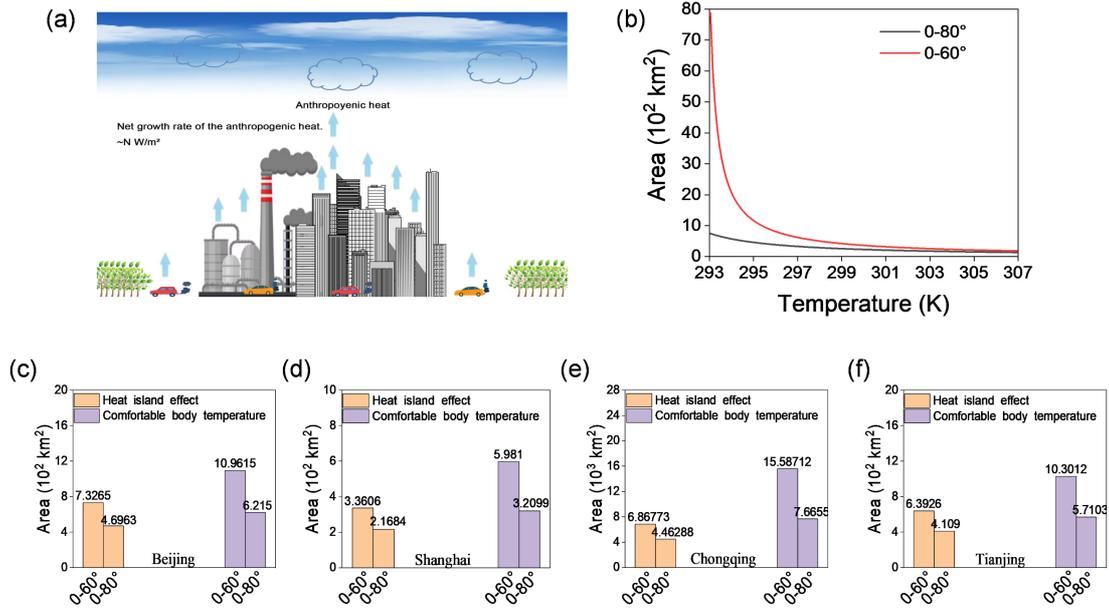

Fig. 5 (a) Schematic diagram of the urban heat island effect. The ~N W/m$^2$ represents the net growth rate of the anthropogenic heat. (b) The calculated deployment area to reduce the temperature by our metasurface radiative cooler with emission angle of 80 ° and 60 °, respectively. (c-f) The calculated area for eliminating the heat island effect and achieving a comfortable body temperature during summertime for China's four big cities: Beijing, Shanghai, Chongqing, and Tianjin.

In addition, as the understanding of radiative cooling deepens and micro/nano fabrication technology develops, complex photonic structures with optimized thermal emission in every wavelength channel and every emission angle across the entire atmospheric transparency window will become necessary to access the limitation of this technology. Furthermore, intelligent radiative cooling with temperature-adaptive capabilities or active control with ultralow energy consumption requires artificially designing dynamic photonic materials and structures. In this context, machine learning methods can undoubtedly open a new pathway and possibility.

**Methods**

**Thermal emitter fabrication.** A Si wafer was firstly cleaned by acetone, isopropanol, and deionized water. Then, plasmon enhanced chemical vapor deposition system was used to deposit the $SiO_2$ and $Si_3N_4$ on top of the Si wafer. A pre-designed hard mask layer was fabricated by sputtering nickel on top of the dielectric layer followed by the standard photolithography and ion beam etching. Then dry etching was employed to etch the $SiO_2$ and $Si_3N_4$ layer. After this, the nickel mask layer was removed by chemical reactions. At last, an optically thick silver layer was sputtered at the back side of the Si wafer with an adhesion layer of Ti.

**Solar reflector fabrication.** The porous PE films were fabricated via phase-inversion-based method. In brief, PE powder, paraffin oil, butylated hydroxyto-luene and other additives were stirred at 150 °C for 3 h to produce a homogenous solution. The heated solution was pushed through a sheet die to make a gel-like film and stretched uniaxially. The as-formed gel was extracted with cyclohexane several times and dried to obtain the porous PE films.

**Structure characterization.** The electron scanning microscope (Zeiss Gemini SEM 300) was used to observe morphologies of the metasurfaces and the distribution of elements in the hybrid elastomer was examined by energy dispersive spectroscope (EDS) conducted on SEM. The optical reflectance of thermal emitter in the ultraviolet, visible and near-infrared (0.3-2.5 μm) wavelength ranges were separately measured using an ultraviolet-visible-near-infrared (UV-Vis-NIR) spectrophotometer (Hitachi, U-4100, Japan) equipped with a deuterium lamp for UV region, tungsten-halogen lamp for Vis, NIR range and a polytetrafluoroethylene integrating sphere. The thermal emission spectra in the mid-infrared wavelength ranges (3-16 μm) were characterized in an FTIR spectrometer (INVENIOR, Bruker) equipped with a deuterated triglycine sulfate crystal detector RT-DTGS, a gold integrating sphereA562 Integrating Sphere, Bruker and KBr beam splitter.

**Rooftop test.** The radiative cooling performance of the sample was tested in January 12[th] and July 6[th] 2022 in Shanghai (East Coast of China, 31°18 22 "N, 121°30 '17" E) under a clear sky. A thermal box was designed with insulation foam covered by a layer of reflective foil, while transparent PE was used to seal the upper part to minimize conduction and convective heat exchange. A relative humidity (RH) data logger (GSP-8, Elitech, Corp, China) with an accuracy of ± 0.1% RH was placed near the thermal box to measure the relative air humidity. A thermocouple (SA1-T-72) temperature detector was mounted on the back of the sample to measure the real-time temperature, and another thermocouple was placed in internal air of the same thermal box (without the sample) to measure the ambient temperature. The solar irradiation incidence on the samples was measured by solar radiometer with an accuracy of ± 0.3% (MS-410, EKO). Two thermocouples, the solar radiometer, were connected to the data recorder (OM-CP-OCTPRO, OMEGA) to record the data. The wind speed around our thermal boxes was measured using a digital anemometer with an accuracy of ± 2.5% (AS856, Smart Sensor Corp, China). Wind speed and temperature were automatically tracked every 30 seconds, solar illumination was recorded every 10 minutes, and humidity was recorded every 10 seconds.

**SUPPLEMENTAL INFORMATION**
Supplementary information can be found online at XXXX.

**ACKNOWLEDGMENTS**
Y.Z. acknowledges the support by the National Natural Science Foundation of China (NSFC) (Grant No. 62175154), the Shanghai Pujiang Program (20PJ1411900), the Shanghai Science and Technology Program (21ZR1445500) and the Program for Professor of Special Appointment (Eastern Scholar) at Shanghai Institutions of Higher Learning. T.W. acknowledges the support by the Shanghai Yangfan Program (22YF1430200). We thank Dr Jingdong Yang and Dr Tianhua Feng for discussion.

**AUTHOR CONTRIBUTIONS**
Y.Z. and M.G. conceived the idea and supervised the research. Y.Z. and Y.C. designed and optimized the metasurface structure by machine learning inverse design. T.W. fabricated the solar reflector and set up the rooftop test. Y.C., T.W and Y.Z. performed the rooftop test. Y.Z. wrote the paper. T.W. and M.G. revised the paper. All authors commented on the manuscript.

**DECLARATION OF INTERESTS**
The authors declare no competing interests.